\documentclass[sigconf]{acmart}
\AtBeginDocument{%
  \providecommand\BibTeX{{%
    \normalfont B\kern-0.5em{\scshape i\kern-0.25em b}\kern-0.8em\TeX}}}

\setcopyright{rightsretained}
\copyrightyear{}
\acmYear{}
\acmDOI{XXX}

\acmConference[]{}{}{}
\acmBooktitle{}
\acmPrice{}
\acmISBN{}

\begin{document}

\title{NationalMood: Large-scale Estimation of People's Mood from Web Search Query and Mobile Sensor Data}

\author{Tadashi Okoshi}
\affiliation{%
  \institution{Keio University}
  \city{Fujisawa}
  \country{Japan}
}
\email{slash@sfc.keio.ac.jp}

\author{Wataru Sasaki}
\affiliation{%
  \institution{Keio University}
  \city{Fujisawa}
  \country{Japan}
}
\email{wataruew@sfc.keio.ac.jp}

\author{Hiroshi Kawane}
\affiliation{%
  \institution{Yahoo Japan Corporation}
  \city{Tokyo}
  \country{Japan}
}
\email{hkawane@yahoo-corp.jp}

\author{Kota Tsubouchi}
\affiliation{%
  \institution{Yahoo Japan Corporation}
  \city{Tokyo}
  \country{Japan}
}
\email{ktsubouc@yahoo-corp.jp}

\renewcommand{\shortauthors}{Okoshi and Sasaki, et al.}

\definecolor{rev}{rgb}{0.0, 0.0, 0.0}

\begin{abstract}
The ability to estimate current affective statuses of web users has considerable potential towards the realization of user-centric opportune services. However, determining the type of data to be used for such estimation as well as collecting the ground truth of such affective statuses are difficult in the real world situation. We propose a novel way of such estimation based on a combinational use of user's web search queries and mobile sensor data. Our large-scale data analysis with about 11,000,000 users and 100 recent advertisement log revealed (1) the existence of certain class of  advertisement to which mood-status-based delivery would be significantly effective, (2) that our ``National Mood Score'' shows the ups and downs of people's moods in COVID-19 pandemic that inversely correlated to the number of patients, as well as the weekly mood rhythm of people.
\end{abstract}



\maketitle

\section{Introduction}
The ability to determine current affective statuses of users has considerable potential to enable the provision of user-centric opportune services tailored to specific user statuses. Web services, for example, can be improved by adapting various types of parameters such as the presentation timings, presentation tone, content, as well as content modality. When Alice has an emotionally negative status, the news web service can highlight some interesting news or present advertisements that can possibly cheer her up, thus attempting to align itself with her emotions.

However, determining affective statuses of web users outside a controlled in-lab configuration, particularly in real-world situations, is a difficult task. 
The first problem is on {\bf the type of data from which the affective status of a user can be estimated}. 
Typically, sensing and determining the emotional state of a person require psycho-physiological data such as heart rate (HR)~\cite{mulder1992measurement}, HRV, electrocardiogram (ECG), and electroencephalogram (EEG) data~\cite{ryu2005evaluation,wilson2002analysis}. 
However, the collection of such data in real-world conditions of mobile web users is not feasible owing to the low penetration rate of such sensors in the society, additional burden on users to use such devices, and lack of social acceptance to the collection of such data.
The second problem is on {\bf how the ground truth label on the affective status of a user can be collected}. User annotation is widely used during the data-collection phase. 
However, this approach is also not always effective owing to multiple possible causes such as the following: (1) the users may find it cumbersome to answer repeated questions, and (2) the users may forget to answer the questionnaire. 

\begin{figure*}[t!]
\begin{center}
\includegraphics[width=1.0\linewidth]{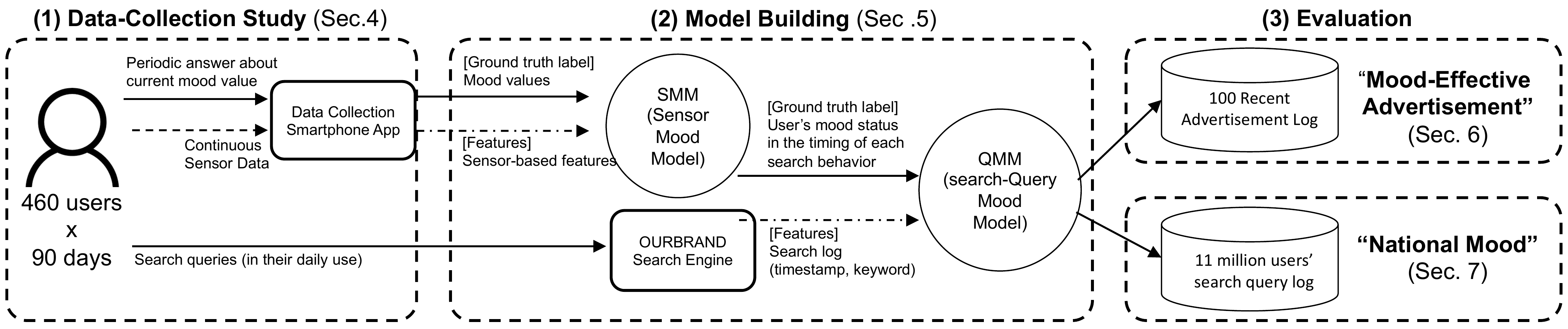}
\vspace{-0.7cm}
\caption{Framework of this research}
\vspace{-0.3cm}
\label{fig:researchframework}
\end{center}
\end{figure*}   

In this paper, as the first contribution, we show that we can estimate the web users' affective status (concretely, ``mood'') in such a condition, based on a novel combinational use of their web search queries and mobile sensor data. 
To address the first problem, we target users' queries input to the web search engine as an easy-to-collect and noninvasive proxy feature to explain their mood states, focusing on the fact that almost all the internet users regularly use search engines in their daily lives. Because web services typically store the historical log of users' search queries at the server side, its use can be started today without having to wait for the wide-spread adoption of new types of sensors. 
To address the second problem, we use a novel two-step mood classification with different types of models, namely the ``Sensor Mood Model (SMM)'' and ``search-Query Mood Model (QMM).'' 

Figure~\ref{fig:researchframework} illustrates our research structure.
(1) First, we conduct a preliminary data-collection study with 460 participants for 90 days to collect their continuous sensor data from their smartphones; periodic subjective evaluation of their mood as ground truth annotation was also performed. 
(2) Next, we build our first model ``SMM'' that estimates the participant's mood statuses from specific temporal frames in which both sensor data and the user annotation were successfully collected. 
With the built SSM, we can estimate each participant's mood status for all the time frames during the data-collection study period. 
Then, by combining the web search logs of the 460 participants during the study period and mood status (based on both the users' original annotation and SMM's outputs), we create our second model ``QMM'', which estimates the mood of a user from their search query data.

As our second contribution, we also show that the use of these mood statuses in a web service (with more than 80 million users) through the introduction of multiple evaluation parameters, including ``Mood-Effective Advertisement'' and ``National Mood,'' has significant potential.  
To the best of our knowledge, this research is the first to reveal this possibility. 

First, by analyzing the existing server-side log data stored in our web service, we investigate the relationship between multiple advertisement contents that we display to the users and the responses of the users based on their derived moods.
By analyzing the recent logs of 100 advertisement projects, including when and who viewed the advertisements and whether they clicked on them, we identified the advertisement for which mood-based delivery will be effective. We determined a certain class of advertisements where the mood tendencies of users who clicked was more positive or negative than a random distribution (Section 6).

Second, we examined the value of calculating the mood score in nation-wide by using the proposed method in this research. We calculated ``the daily national mood score'', the average mood score of about 11,000,000 users in Japan on a daily basis, and saw how the score changes over time. Interestingly, we found that the score based on our proposed algorithm shows the weekly rhythm of people's mood (that drops every 1st working day of the week and increases again every weekend) (Section 7.3) and the longer-term trace of people's mood in the COVID-19 pandemic period in year 2020 that inversely correlated to the number of COVID-19 patients (Section 7.4). 

The remainder of this paper is organized as follows. 
Section~\ref{sec:RelatedWork} discusses related work.
Section~\ref{sec:Proposition} describes our novel Yahoo! JAPAN  Emotion framework.  
Section~\ref{sec:DataCollection} details data-collection study conducted for 90 days with 460 users. Section~\ref{sec:ModelBuilding} describes model building from the collected data. 
Section~\ref{sec:EvaluatoinAdvertisement} shows the results on ``Mood-Effective Advertisement'' analysis. Section~\ref{sec:EvaluatoinNation} shows our findings from our ``National Mood Score'' analysis.
Section~\ref{sec:fw} discusses our further work.
Section~\ref{sec:Conclusion} concludes this paper.


\section{Related Work}
\label{sec:RelatedWork}

Extensive studies on emotion began to be conducted in the 19th century, 
with a well-known study conducted by Darwin~\cite{ekman2006darwin}. 
Darwin says that emotion is a product of evolution and that emotions induce actions favorable to survival~\cite{darwin1998expression}.
Numerous emotional modalities and their respective physiological responses have been studied~\cite{royet2000emotional, craig2002you}.
Emotional states are known to affect cognitive and athletic abilities and are reported to affect both human-human and human-machine interactions.
Picard generalized this research field as ``affective  computing''~\cite{picard1997affective}. 
Many studies and systems have been proposed to detect and utilize the emotions of users~\cite{chang2011s, de2003real, khan2013towards, likamwa2013moodscope, scheirer2002frustrating} in this field.
Several methods of determining user emotion have been proposed, which focus on physical characteristics~\cite{kwon2007emotion,Goncalves:2014:PTD:2632048.2636067}, text data~\cite{Mohammad:2010:EEC:1860631.1860635,wang2012harnessing}.
In our research, we focus on the mood status of users. 
Mood is related but different from emotion in several aspects~\cite{beedie2005distinctions}.
Mood is usually tend to last longer than emotion and is usually a cumulative reaction 
while emotion is a more spontaneous reaction or emotion caused by a particular event. 

In recent years, various research have been conducted to estimate the mood state of the user by focusing on the sensors on smartphones.
Smartphones are equipped with multiple sensors and can collect a wide variety of data, 
such as acceleration, rotation, location, and network connectivity. 
Also, smartphone can now be considered indispensable to our lives with, for example, more than 95\% 
penetration rate in Japan.
In addition, smartphone-based sensing do not require additional devices in sensing, 
such as those for measuring heart rate, EEG or ECG, that may have additional burden on the users. 
Owing to the rapid development and spread of smartphones, various studies have been conducted 
on the recognition and estimation of the mood status of a smartphone user.
Most studies have constructed a classification model that determines moods from 
the user's contextual data obtained from the smartphone sensor data 
and self-reporting annotation by the user~\cite{ma2012daily}.
MoodScope~\cite{likamwa2013moodscope} investigated the effects of the user context 
on the mood of a user based on the smartphone sensor data.
In addition to emotion and mood, various types of internal statuses of the users, 
such as ``interruptibility''~\cite{pejovic2014interruptme,okoshi2019real}, have been recognized and estimated from the smartphone data.
Other types of sensing modality for emotion estimation is facial expressions in the image data. 
Such research has been widely conducted~\cite{fasel2003automatic}, mainly by using Facial Action Coding System~\cite{ekman1997face}. And some research on the smartphone~\cite{suk2015real} platform have been also performed recently.
In contrast to those previous works, our research highlights (1) its novelty in the combined use of smartphone sensor data and web search queries, and also a large-scale data collection study and data analysis.

Focusing on the data on the web, researchers are currently working on 
estimating the emotional states of users 
by analyzing text data on social networks such as Twitter~\cite{wang2012harnessing, bollen2011modeling} and Facebook~\cite{kramer2012spread, kramer2014experimental}. 
These social network text data may contain sentences containing the affective information of users.
In this research, we focus on the web search queries to estimate the users' mood status since they are easy-to-collect noninvasive type of data. 
In reality, it is difficult to estimate the affective state directly by using such queries 
since most of them comprise a few words and nouns. 
Again, our novel approach focuses on a combination with different types of mood estimation models from smartphone sensors and search queries.

\section{Yahoo! JAPAN  Affective Service Framework}
\label{sec:Proposition}

\begin{figure}[!t]
\begin{center}
\includegraphics[width=0.95\linewidth]{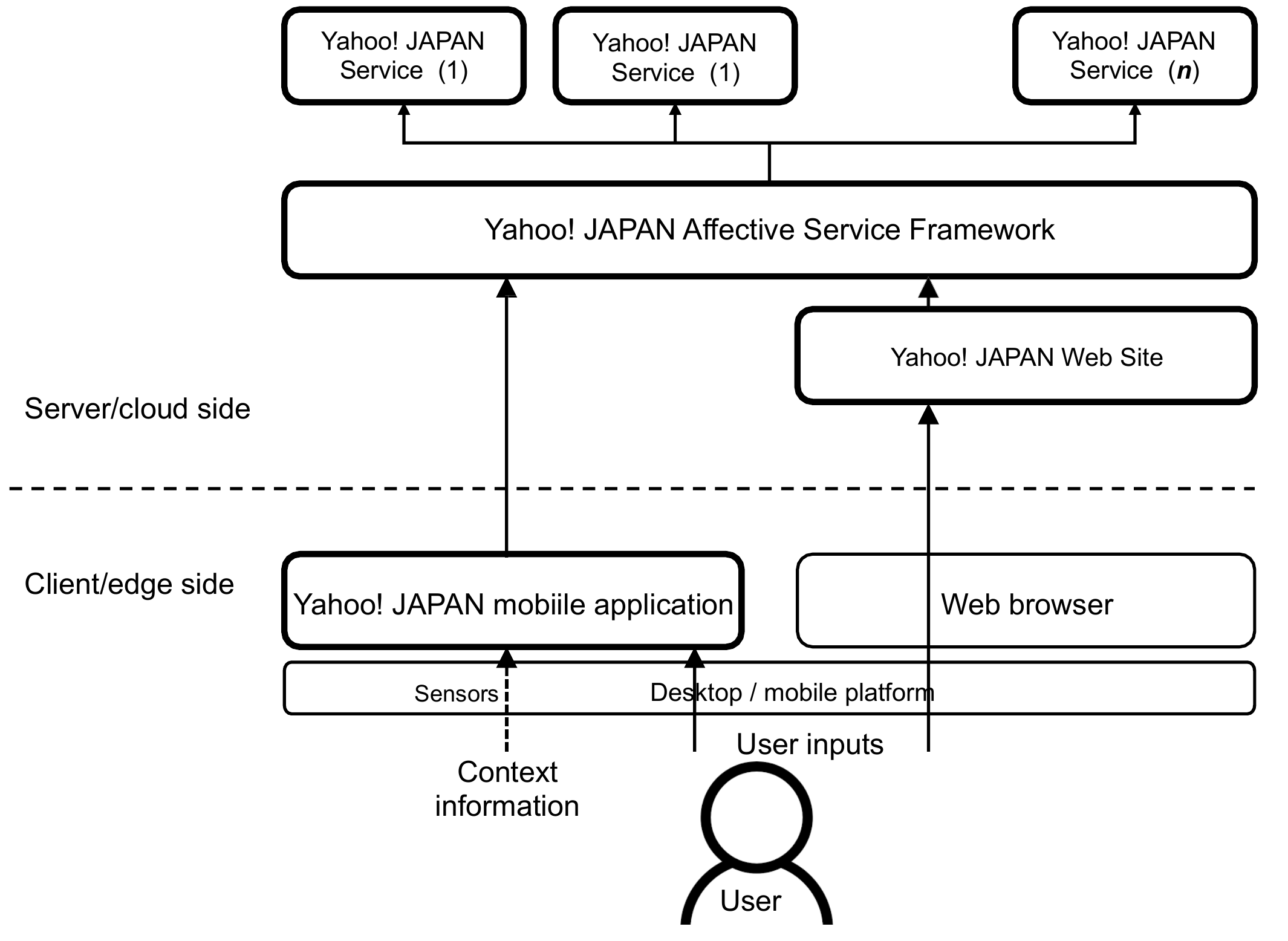}
\vspace{-0.2cm}
\caption{Yahoo! JAPAN  Affective Service Framework}
\vspace{-0.1cm}
\label{fig:yahoo_emo_framework}
\end{center}
\end{figure}   

Figure~\ref{fig:yahoo_emo_framework} shows the conceptual view ``Yahoo! JAPAN  Affective Service Framework'' being developed on Yahoo! JAPAN  web service. 
Yahoo! JAPAN  has a widely-known and widely-used web site in Japanese market, with more than 70 Yahoo! JAPAN -branded services such as ``search'', ``news'', ``shopping'', ``auction'', ``movie'' and ``weather'' and with the total number of 80 million registered users. (Note that Japan's national population is about 126 million.)
In addition to the conventional web pages optimized both for PC and mobile devices, Yahoo! JAPAN  has its own smartphone application both on iOS and Android platform. 
More than 60 million users are using these applications, making them one of the most popular smartphone applications in this country.

Our view around Yahoo! JAPAN Affective Service Framework is as follows. At the server side, each of more than 70 Yahoo! JAPAN  services are logging users’ usage including page view and its duration, clicks, and various types of inputs including web search keywords as the most major example of such kind. In addition, at the users’ client side, our smartphone application can opportunistically collect various types of sensor data on the users’ smartphone, given the users’ permission.
Hence, with such multiple types of input data both from the user's client and server sides, we can opportunistically estimates the user's affective status by using the machine learning techniques, and share the results with diverse Yahoo! JAPAN  services on top of the that so that those services realize service-specific affective-aware adaptation and optimization. 

\section{Data Collection Study}
\label{sec:DataCollection}
Towards the realization of such framework above, firstly collecting data and the user's subjective mood evaluation from the real-world is inevitable. Thus, we first conducted a data-collection study with 460 users for 90 days. We collected continuous data from various types of smartphone sensors as well as the user's subjective mood evaluation (up to 6 times a day) as the ground truth label. 

\begin{figure}[!t]
\begin{center}
\includegraphics[width=0.95\linewidth]{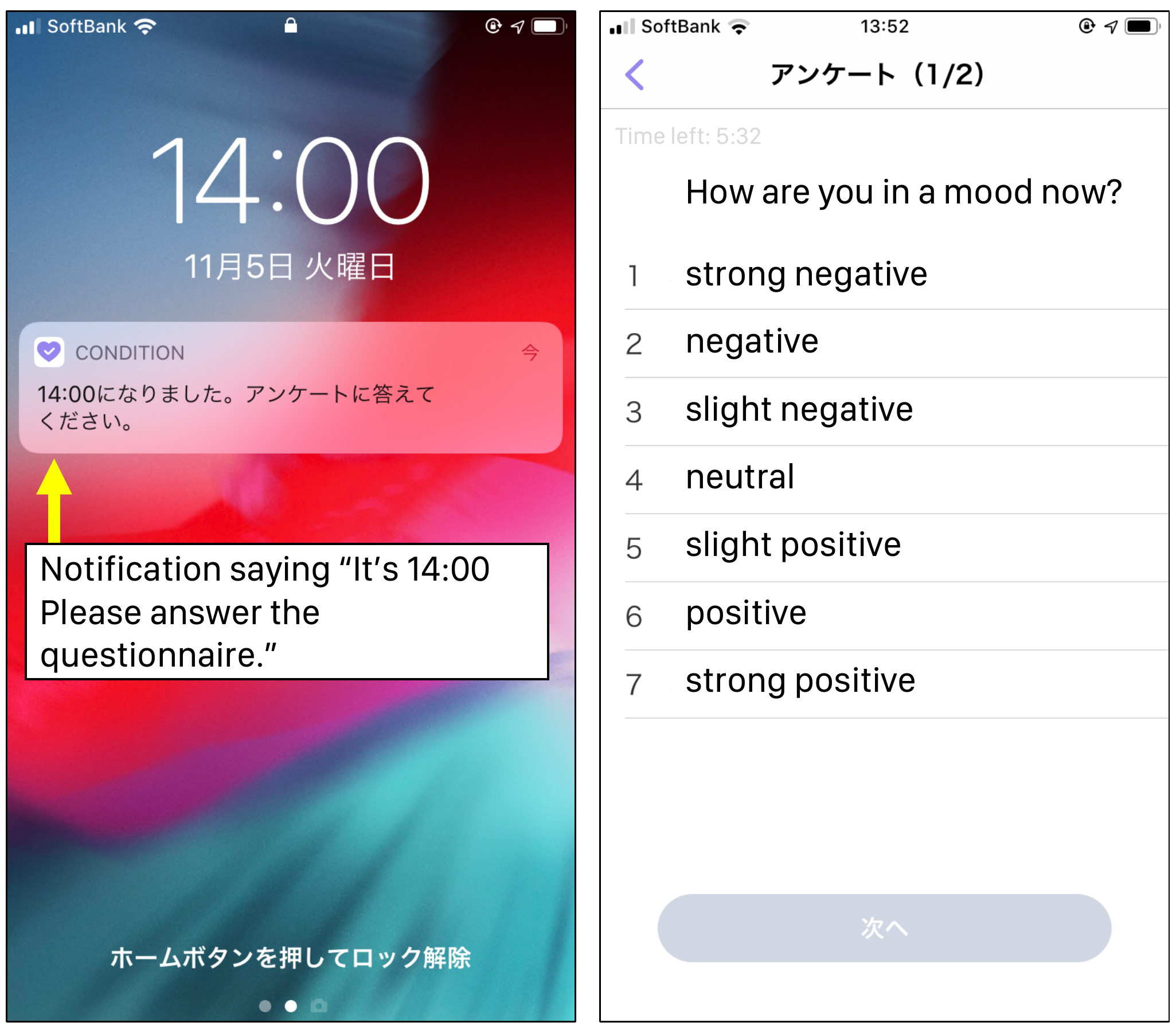}
\vspace{-0.2cm}
\caption{Screenshot of Data Collection Application}
\vspace{-0.2cm}
\label{fig:app_screenshot}
\end{center}
\end{figure}   

\subsection{Participants}
For the study, participants were recruited through an external agency. 
The recruitment criteria were as follows: (1) the age should be in the range of 18-59 years,  (2) must own an active Yahoo! JAPAN  registered account, (3) must have the ability to use the Yahoo! JAPAN  search functionality once or more times per week and should have performed a search at least once in the last month, (4) must own and use an Apple iOS smartphone as a private primary phone in their daily life, and (5) must be a user of an iPhone 7 or later and iOS version 12 or above. The participants were informed that this study was ``an experiment about your condition'' during the recruitment process. 

A total of 460 users, consisting of university students, staff members, and research engineers aged between 19 and 54 years (average:36.92), were recruited. 

\subsection{Duration}
The study was conducted for 90 days, from November 1, 2019 to January 29, 2020. 

\subsection{Experimental Setup}
We developed a dedicated smartphone application in this study, as illustrated in Figure~\ref{fig:app_screenshot}. The application was developed for the iOS platform owing to several reasons. First, the market share of iOS is higher than that of Android in the Japanese market; thus, the recruitment of participants is easier. Second, the number of iPhone models (such as iPhone 7, 7Plus, 8, 8Plus, X, and 11) of Apple is lower than that of Android phones (hundreds of models by dozens of manufactures with different OS-level optimization in power management, sensing, etc.). Thus, we can easily test the application with such phone models to achieve higher execution stability. Third, thanks to the iOS AWARE Framework~\cite{ferreira2015aware,nishiyama2020ios}, we can implement and deploy an application that can continuously collect various sensor data in spite of the fact that iOS is a rather strict environment as a sensing platform than Android.

Once the application is installed on the smartphone of a user, it continuously collects multiple types of data from the embedded sensors of the phone, as detailed in Table~\ref{session_data}, and periodically uploads the data to the server. 

The application can also issue a notification (as shown on the left-hand side of  Figure~\ref{fig:app_screenshot}) at the timings configured by the developer. Once the user responds to the notification, the application opens a questionnaire, where the user can report his current mood status on a 7-level Likert scale (1. strongly negative, 2. negative, 3. moderately negative, 4. neutral, 5. moderately positive, 6. positive, and 7. strongly positive).

\begin{table}[t]
  \begin{center}
      \caption{Sensor Type and Sensing Frequency}
   \small 
    \begin{tabular}{cc} \toprule
Sensor Type & Sensing Frequency\\ \midrule
Accelerometer & 10Hz \\
Barometer & 1Hz \\
Battery status & 1Hz \\
Location & 1/180Hz\\
Network type & (changed event-based)\\
Weather (from OpenWeather) & 1/60Hz\\
Screen status(On/Off) & (changed event-based)\\
\bottomrule
    \end{tabular}
    \label{session_data}
  \end{center}
\end{table}

%
%
%
%
%
%
%
%

\subsection{Experimental Procedure}
Our experimental procedure consisted of the following three parts:  
(1) Each participant had a meeting with a study researcher at the beginning of the study and received basic information and instructions about the study, which was followed by the signing of a consent form. Next, the participants were asked to install and launch our software on their smartphones.
They were asked to grant the following permissions to the application: (1) mic, (2) push notification, (3) motion and fitness activity, and (4) location (configured as ``always'') data sensing feature of the iOS platform.

(2) After the initial meeting, the 30-day study period started. 
During this period, a push notification appeared six times (at 8:00 AM, 10:00 AM, noon, 2:00 PM, 4:00 PM, and 6:00 PM) every day. Each participant was asked to proceed with the survey within 2 hours after the delivery of each notification. 
When the participant opened a notification, the application screen (Figure~\ref{fig:app_screenshot}) appeared and asked the affective mood of the user on the 7-level Likert scale. The participant selected the status, and after a confirmation prompt, the answer was submitted to the server.

\subsection{Reward}
We created an instant point reward system in the application. Each participant scored 0, 20, 30, or 40 points for 0-3, 4, 5, or 6 answers, respectively, in a day. The earned reward points were accumulated throughout the study period. When a participant reached the configured minimum total reward points, i.e., 1500 points (by answering 4 answers every day for 75 days or 6 answers every day for 38 days), they received a payment. They received an additional payment when they exceeded 2000 points. 



\section{Model Building}
\label{sec:ModelBuilding}
This section describes our model building of two different models, Search Mood Model (SMM) and search-Query Mood Model (QMM) respectively.
As introduced and illustrated in Figure 1, SMM classifies user's mood from a set of features computed from sensor data. Using the built SMM model, having more training data, we build QMM that estimates user's mood from search queries.




\subsection{Sensor Mood Model (SMM)} 
For building Search Mood Model (SMM), we follow an approach with time-frame-based feature extraction of the time-series sensor data and their classification which are widely used in activity recognition area~\cite{bao2004activity}.

\begin{table}[t]
  \begin{center}
      \caption{Extracted Features}
\resizebox{1.0\columnwidth}{!}{
    \begin{tabular}{ccl} \toprule
{\bf Sensor Type} & {\bf Number of Features} & {\bf Representative Features}\\ \midrule
Accelerometer & 23 & 
\begin{tabular}{l}
(mean, std, median, min, max) magnitude \\ mean (each axis) \\ variance (each axis) \\ variance (each axis) \\ skew (each axis) \\ kurtosis(each axis) \\ correlation (xy, yz, zx) \\ covariance (xy, yz, zx) \\
\end{tabular}\\ \midrule
Barometer & 5 &
\begin{tabular}{l}(mean, std, median, min, max) magnitude \end{tabular}\\ \midrule
Battery status & 7 &
\begin{tabular}{l}
(mean, std, median, min, max) battery level \\
number of charge times \\
length of charge minutes \\
\end{tabular}\\ \midrule
Location & 12 & 
\begin{tabular}{l}
location entropy \\
number of location transitions\\
moving time percent\\
\end{tabular}\\ \midrule
Network type & 5 &
\begin{tabular}{l}
number of WiFi connectivity established\\
number of mobile connectivity established\\
most frequent network type \\
rate of WiFi \\
rate of mobile network\\
\end{tabular}\\ \midrule
\begin{tabular}{c}
Weather\\ (from OpenWeather) 
\end{tabular}
& 50 &
\begin{tabular}{l}
weather type \\
(mean, std, median, min, max) temperature \\
(mean, std, median, min, max) humidity \\
\end{tabular}\\ \midrule
Screen status(On/Off) & 11 & 
\begin{tabular}{l}
number of unlocks (per minute) \\
number of interaction (per minute)\\
\end{tabular}\\ \bottomrule
    \end{tabular}}
    \label{tbl:features}
  \end{center}
\end{table}

\subsubsection{Feature Extraction}
First, we extracted features of each 3-hour time window from the collected raw sensor data obtained from our study with 460 users for 90 days.
The features were extracted for each sensor type. The types of extracted feature are different depending on the sensor types. 
Table~\ref{tbl:features} summarizes the number of features extracted along with several representative feature types.

\subsubsection{Model Building}
Then we constructed a supervised machine learning model from frames base on the extracted feature vector and a self-reported mood status as its ground truth label.
In this model building, we treat the self-reported mood status as a three-class classification problem.
The collected mood answers (originally in the 7-level Likert scale) were assigned to 3 different labels, $-1$ for ``strongly negative'', ``negative'', and ``moderately negative'', $0$ for ``neutral'', and $+1$ for ``strongly positive'', ``positive'' and ``moderately positive.''
We chose Random Forest~\cite{RandomForest} for the machine learning algorithm which revealed the best classification performance compared with others.

\subsubsection{Cross-validation Performance}
According to a 10-fold cross-validation for the built model, the model performs an accuracy of 66.6 \%.
Table~\ref{tbl:mood_perfomance} shows the results of the overall performance evaluation.

\begin{table}[t]
  \begin{center}
      \caption{Performance of Search Mood Model (SMM)}
      \resizebox{0.6\columnwidth}{!}{
    \begin{tabular}{cccc} \toprule
    Label & Precision & Recall & F1-score \\ \midrule
    -1 & 0.23 & 0.36 & 0.28 \\ 
    0 & 0.41 & 0.54 & 0.46 \\
    1 & 0.85 & 0.73 & 0.78 \\ \midrule
    avg & 0.72 & 0.67 & 0.69 \\ \bottomrule
    \end{tabular}
    }
    \label{tbl:mood_perfomance}
  \end{center}
\end{table}

\subsection{Query Mood Model (QMM)} 
QMM is a model that examines the relationship between a user's web search query and the user's mood score during the search behavior. After its training, it classifies the user's mood score from their search query data. Here, there can be two different types of mood score, namely (a) the scores answered by the data collection participants with questionnaire and (b) the scores estimated by SMM based on the collected sensor data. In this section, we explain the concept of QMM and how additional use of (b) increases the performance of QMM.

\subsubsection{Model Building}
For each fixed time range (what we call ``session''), QMM gets trained from the data of a user's mood score and search behavior during the session. 

To investigate the validity of the trained QMMs qualitatively, we employed logistic regression, which is a typical example of a ``white box'' model with high model interpretability, as a model.
Note that it is not necessary to specify the logistic regression as training scheme in actual operation; we believe that non-linear SVMs and decision tree-based regressions such as Xgboost (that are specialized for performance) are also effective.

One session was defined as one record, and training was performed in the following regression equation.

\begin{eqnarray}
    y = \theta_0 + \theta_1 x_1 + \theta_2 x_2 + \dots + \theta_n x_n ,
\end{eqnarray}

where $y$ is the mood score and $\theta_k$ is the learned weight. $x_k$ is the search query assigned to a feature, with 1 if it was searched in that session and 0 if it was not.
$x_k$ indicates only whether the query is searched or not, not considering the number of searches.

\subsubsection{Combination with SMM}
In this training, sessions that do not have both ``search behavior'' and ``mood score'' data will be treated as ``missing data''. Therefore, mood scores based on the users' raw survey response data have the challenge in their limited number of responses available for training.
In such a situation, our SMM model, along with collected sensor data, is an effective means of increasing the number of mood scores to be used for the training. Since the sensor is always on as long as the smartphone (with our application) is on, SMM can be used to estimate a user's mood virtually for  24/7.

We use a session of 3 hour long. The search queries retrieved in the three separated hours were used as features. Since SMM-based mood scores are available for 24 hours a day, logically the length of each session can be more short (fine-grained). However, if the length is too short, there is a risk that the questionnaire-based mood scores of the comparison method may become too sparse to be learned. 
Hence, the sessions needed to be reasonably long. 
From such discussion, we decided to use the session length of three hours in this study.

\subsubsection{Cross-validation Performance}
We built 2 QMM models, one trained only from the questionnaire answer data, and another one with additional training data based on the SMM outputs. 

For both models, the condition of the performance evaluation is as follows. 
The data were randomly divided into 80\% training data and 20\% evaluation data. 
Considering the effect of randomness, the evaluations were conducted for 10 times. 
The training data were balanced so that the amount of positive and negative data was the same before training. For the evaluation data, we did not perform balancing.

The results are shown in Table~\ref{tab:acc}, which clearly confirms the effectiveness of SMM. 
Compared to our baseline QMM without SMM use, the accuracy increases from 87\% to 94\% in case with additional data brought by SMM. 
The table also shows that the number of training data has been more than doubled by SMM, indicating that the more than doubling of the dataset used for training by SMM has contributed greatly to the significant improvement in prediction accuracy.
From these results we conclude to adopt a QMM trained with SMM in this research, and go on to our evaluation experiments. 

\begin{table}[t]
\caption{Performance of search-Query Mood Model (QMM)}
\resizebox{0.7\columnwidth}{!}{
\begin{tabular}{lcc}
\toprule
\multicolumn{1}{c}{} & \# of data & accuracy \\ \midrule
QMM (with SMM (proposed))  & {\bf 46,318}     & {\bf 94\%}     \\
QMM (without SMM)          & 27,402     & 87\%     \\ \bottomrule
\end{tabular}}
\label{tab:acc}
\end{table}

\section{Evaluation 1: Mood-Effective Advertisement}
\label{sec:EvaluatoinAdvertisement}
To investigate how our mood model works effectively, we evaluated it with our past web advertisement business cases.  
We conducted an experiment on 100 advertisement cases that were actually delivered to our users in the past.
We examined the impression (the number of times the advertisement was exposed to users) and click logs for those 100 ads, and tested offline to see if there were ads that made a difference in whether users clicked or not, depending on their mood state just before their click.

\subsection{Dataset}
The targeted dataset consisted of 100 ads served through Yahoo! JAPAN 's ad service.
To ensure sufficient amount of impression and click volume for the data analysis, the 100 ads were selected from historical ad business data stored in our servers as the most recent 100 projects with a delivery record of at least 14 days.
The targeted ads had at least about 5 billion user impression (view) and about 3 million clicks during the 14 day period in total.
User IDs and timestamps for impressions and clicks on ads are respectively stored in our server's internal storage. 
Therefore, through the user ID in each impression or click log, we can link the user's history of organic search behavior.
Through think link, for each user, their three-hourly mood score can be calculated by using our QMM model. 
Finally, the results of this calculation can be used to analyze whether or not a user in a certain mood state clicked on an ad when they viewed it.

\subsection{Metrics}
In this study, a pairwise method was used to investigate whether there are ads that change the click state of an ad depending on the user's mood score.
First, we organized the logs for the days the ads were served and converted the data into a format of "timestamp, the viewer's user ID, if clicked/or not."

Then, for each day, we randomly select two records from all the records. 
If only one of the two extracted records is a "clicked log", we compare the mood scores of the two records each other. We mark ``positive'' when the clicked user's mood score is higher, and vice versa. 
This pairwise extraction process was performed 100 million times for each day of the log, to determine whether each ad was more likely to be clicked on when the user's mood score was positive or negative.

\subsection{Result}
\subsubsection{Existence of ``Mood-Effective Ads''}
\begin{figure}[t]
\begin{center}
\includegraphics[width=0.95\linewidth]{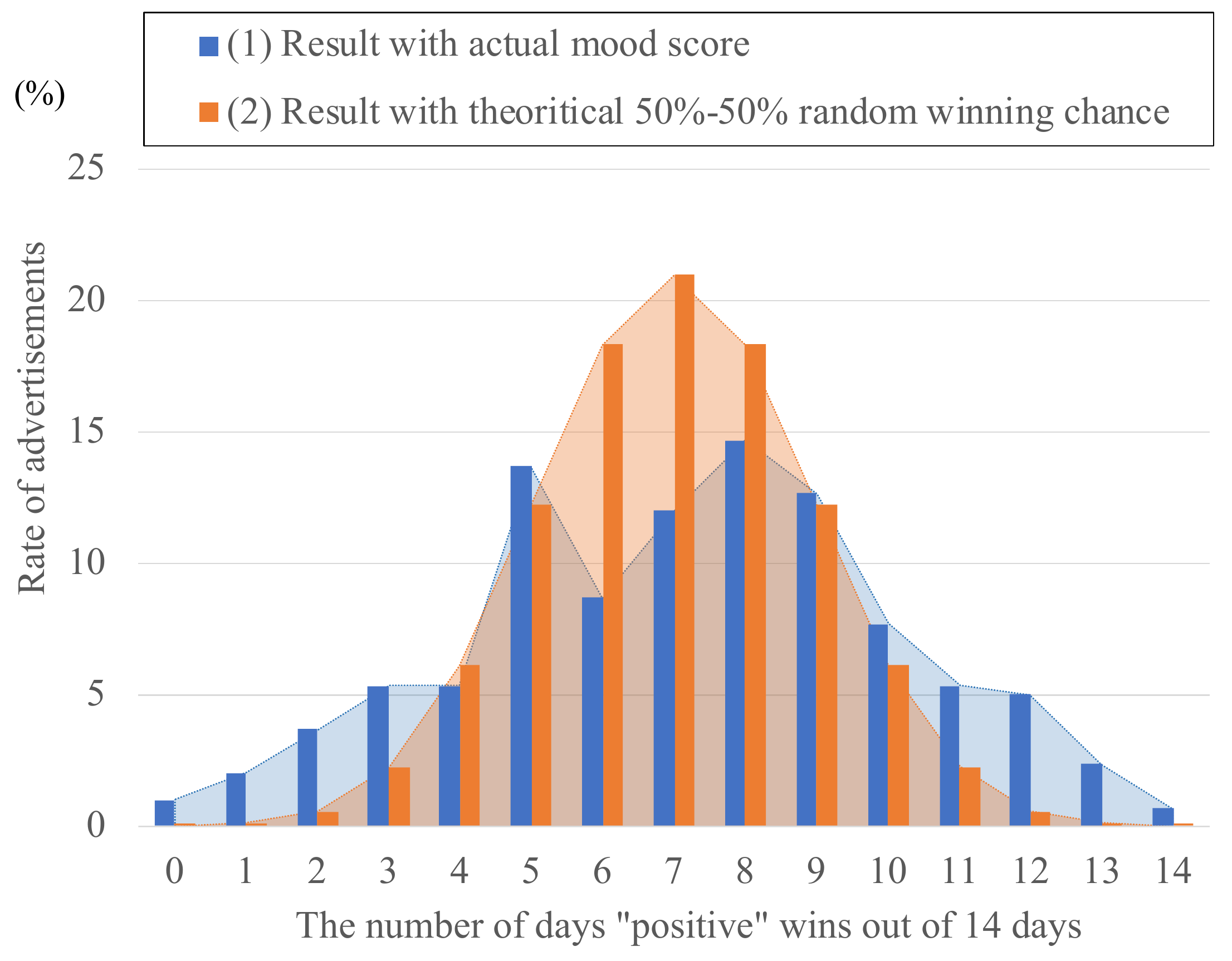}
\caption{Distribution of Advertisements by Positive/Negative Ratio}
\label{fig:res1}
\vspace{-0.2cm}
\end{center}
\end{figure}

We show the result in Figure~\ref{fig:res1}. 
The x-axis is the number of days ``positive'' wins out of 14 days, 
while the y-axis shows the percentage of such advertisements (among 100 ads).
Two lines are depicted in the figure; they are (1) the pairwise comparison result based on the actual estimated mood scores and, and (2) a theoretical line that would be drawn if we assume that the winner in each pairwise comparison attempt was completely random (50\% - 50\%).

Firstly, as we can understand easily, the shapes of the graphs (2) random theoretical values have a convex shape that expands toward the center.
In other words, the line (2) can be said to be the result of a hypothetical experiment of 14 consecutive coin tosses. 
Thus, obviously 14 consecutive win (or lose) out of 14 tries is an very rare event.
(The probability of a positive winning in all 14 days is about 0.01\% when we randomly assign a score. This is the probability of 1 out of every 10,000 ad serving.)
Similarly, the same kind probably with 13 wins (or losses) is about 0.1\% (1 out of 1,000). 

However, a clear difference can be observed when looking at the actual data (1).
The number of ads where the positive wins on all 14 days or loses on all 14 days can be found in about 1\% (1 out of 100) of the ads, if they would be served by the mood score.
Furthermore, the results with ``more than 13 wins or losses'' scored 6 (6\%).
We also can observe that the shape of the line (1) convexity is spread out to look like it has been crushed to the side. 
From these results, we can confirm that there are indeed certain advertisements that are effective for delivery based on the recipients' mood scores.

\begin{figure}[t]
\begin{center}
\includegraphics[width=0.95\linewidth]{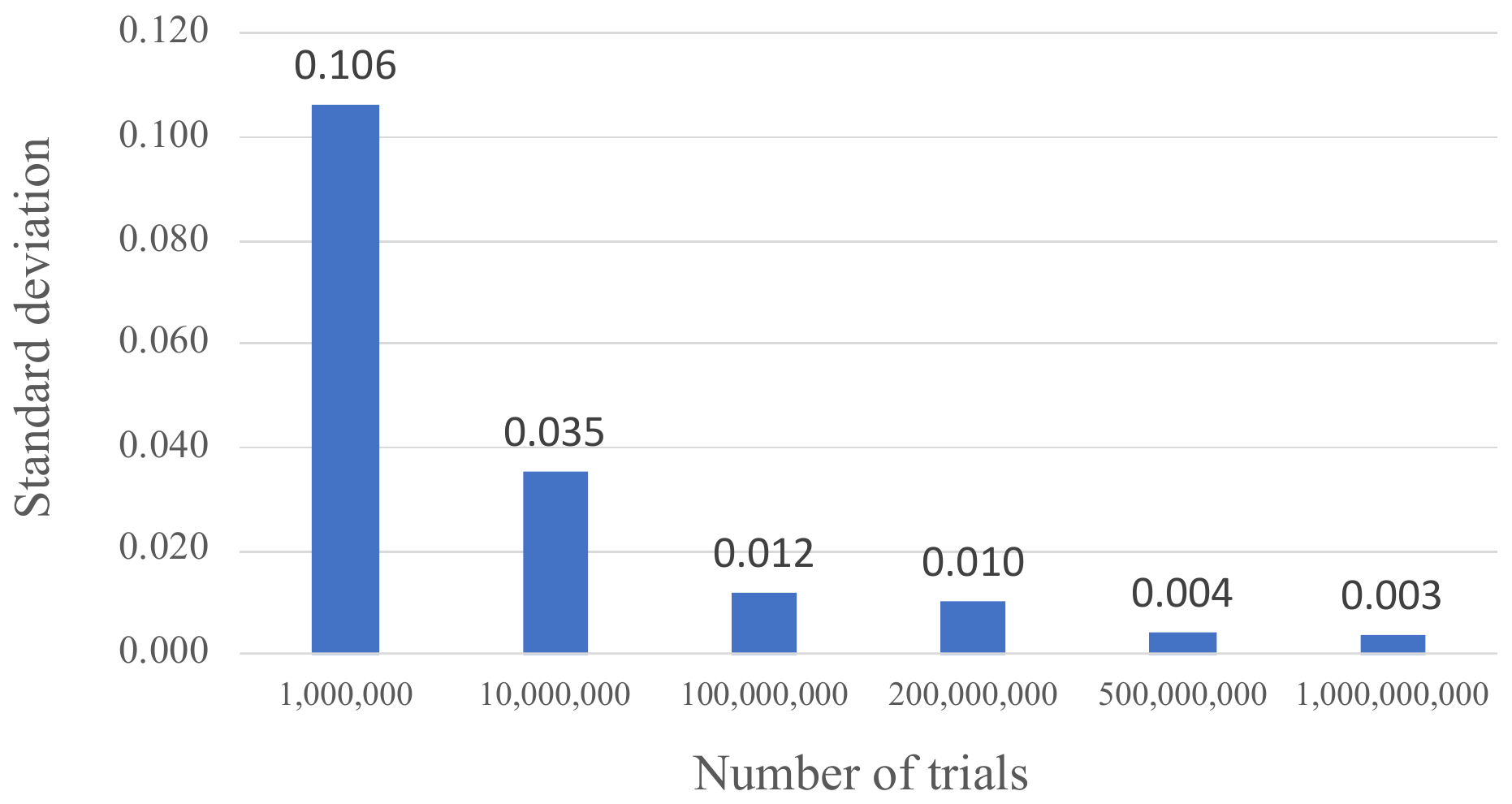}
\caption{Standard Deviation Values with Different Number of Trials}
\label{fig:res1_stdev}
\end{center}
\end{figure} 

For the number of trials ``100 million'', we tried several different numbers and confirmed that the variability was sufficiently small with this number of iteration. 
Figure~\ref{fig:res1_stdev} shows the standard deviation scores for the positive/negative judgment of the ads in case of different number of trials in the pairwise comparison method. 5 ads were arbitrarily selected, the metric was applied, and the standard deviation of the scores of the positive/negative ratio was obtained. As shown in the figure, the score obtained after one million trials has 0.106, which means that the result may increase or decrease by about 10\%. 
On the other hand, with 100 million trials, the standard deviation was about 1.2\%, which means that the score does not change by more than that number of trials. At this level, we determined that the score would be statistically reliable.

\subsubsection{Detailed Ad Content Analysis}
We further examined in detail the ads that were more likely to be clicked on in the same emotional state for more than 13 out of 14 days.
Although we cannot provide examples of actual advertisements delivered due to the confidentiality of the business-related information, we found that the ads that were most likely to be clicked on when the user was in a good mood were those that contained the keywords "free" and "deals," and that many users were considering purchasing a product or service in the future.
On the other hand, the common denominator of advertisements that were more likely to be clicked on when users were in a bad mood was the content of the advertisements to relieve users of their complaints.
It is very interesting to note that there were clear differences between the ad groups that were more likely to be clicked on in good and bad moods.

\section{Evaluation 2: National Mood}
\label{sec:EvaluatoinNation}
The purpose of this experiment is to examine the value of calculating the mood score in nation-wide by using the proposed method.
Regarding the registered users of Yahoo! JAPAN , our internal data on their demographics shows that almost all of them are geographically located in Japan. Therefore, by calculating and averaging the mood scores of all those users on a given day, we can derive a value that we call ``the national mood score'' of Japan on a daily basis.

\subsection{Daily National Mood Score}
In this evaluation, for each day of the given dataset, we (1) computed (individual) daily mood scores of approximately 11,000,000 Yahoo! JAPAN  search users from their search query logs previously acquired and stored, and (2) calculated the average of them. We name this average score as ``Daily National Mood Score''. Then, we (3) examined the changes of this score over time. 
Note that the exact number of the users for the calculation changed day by day, since we targeted Yahoo! JAPAN registered users who used our search engine for at least once, for each particular day.

For our 2 types of evaluation goals, we used two different datasets with different periods and duration, as we present in Section 7.3 and 7.4.

\subsection{Comparative Method}
As described earlier,  one of the challenges in this research is to show the effectiveness of SMM when used inside the QMM training. Thus, we compare (1) QMM with SMM as our proposed method and (2) QMM without SMM as a comparative method in this evaluation. 
Note that all the conditions (algorithm, hyper parameters, split ratio between the data) are the same between these two methods. 

\subsection{Result 1: Weekly Mood Rhythm}

The first case is an analysis of the trend of daily mood score for four weeks. We want to see how national mood score changes within a month, relatively a short term. 
For such analysis, we used a dataset for the period from July 1, 2019 to July 28, 2019. 
We carefully chose this period from the stored log to avoid including major news that could have a significant impact on many users.

\begin{figure}[t]
\begin{center}
\includegraphics[width=1.0\linewidth]{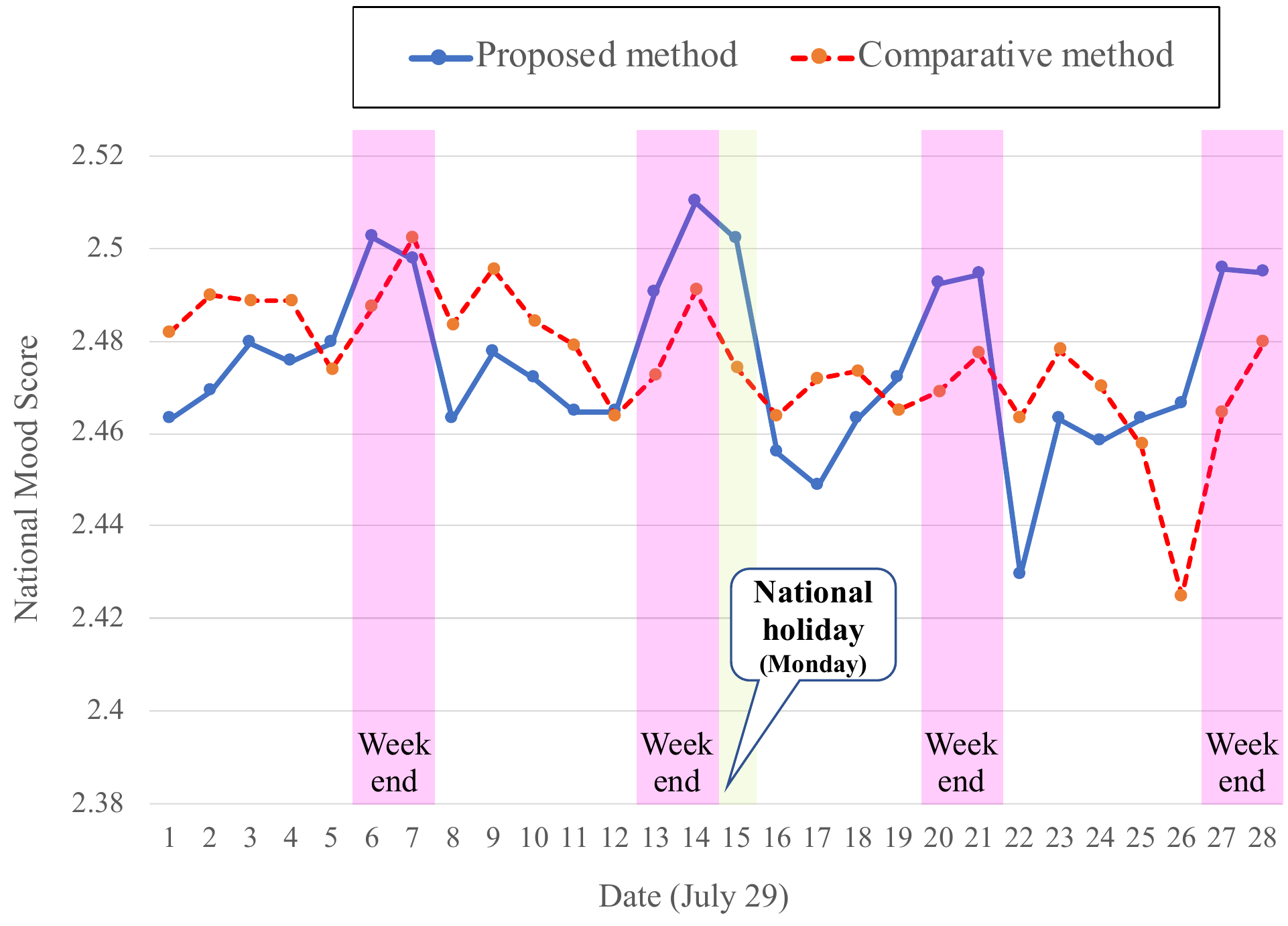}
\caption{Daily National Mood Score for Weekdays and Weekends}
\label{fig:res2}
\end{center}
\end{figure}

Figure~\ref{fig:res2} shows the resulted national mood scores for this period. 
The x-axis represents the date while the y-axis represents the daily score.
A very interesting result we can read in the figure is that the scores tend to be clearly positive on weekends and more negative on Mondays when the workday begins (or Tuesdays when Monday is a holiday on July 15).

Although no ground truth data exists on nation-wide mood, this tendency to feel better later in the week and then worse again on Monday is considered to be a visceral result in a society where many people work Monday through Friday, as there is the term "Blue Monday." 
In fact, we have some facts that can explain this. 
According to a white paper by the Ministry of Health, Labour and Welfare~\cite{JisatsuWP2019}, the highest number of suicides are on Mondays in Japan.
Another survey of 400 men and women, conducted by a company, found that the highest number of respondents in all age groups said they felt most depressed on Mondays~\footnote{\url{https://kyodonewsprwire.jp/release/201803262310}}.
On the other hand, it's easy to imagine the mood being more positive on weekends and holidays.

The proposed method looks successfully expressing the rhythm of the mood change over the weekdays and weekends in an instructive manner, while such rhythm is not very clear in the comparative method. 
In particular, the proposed method is clearly showing the depression on Monday (or Tuesday if Monday was a national holiday). 
From this result, we discuss that the proposed method is better explaining the weekly mood rhythm than the comparative method. 

\subsection{Result 2: COVID-19 and National Mood}
\begin{figure*}[t]
\begin{center}
\includegraphics[width=1.0\linewidth]{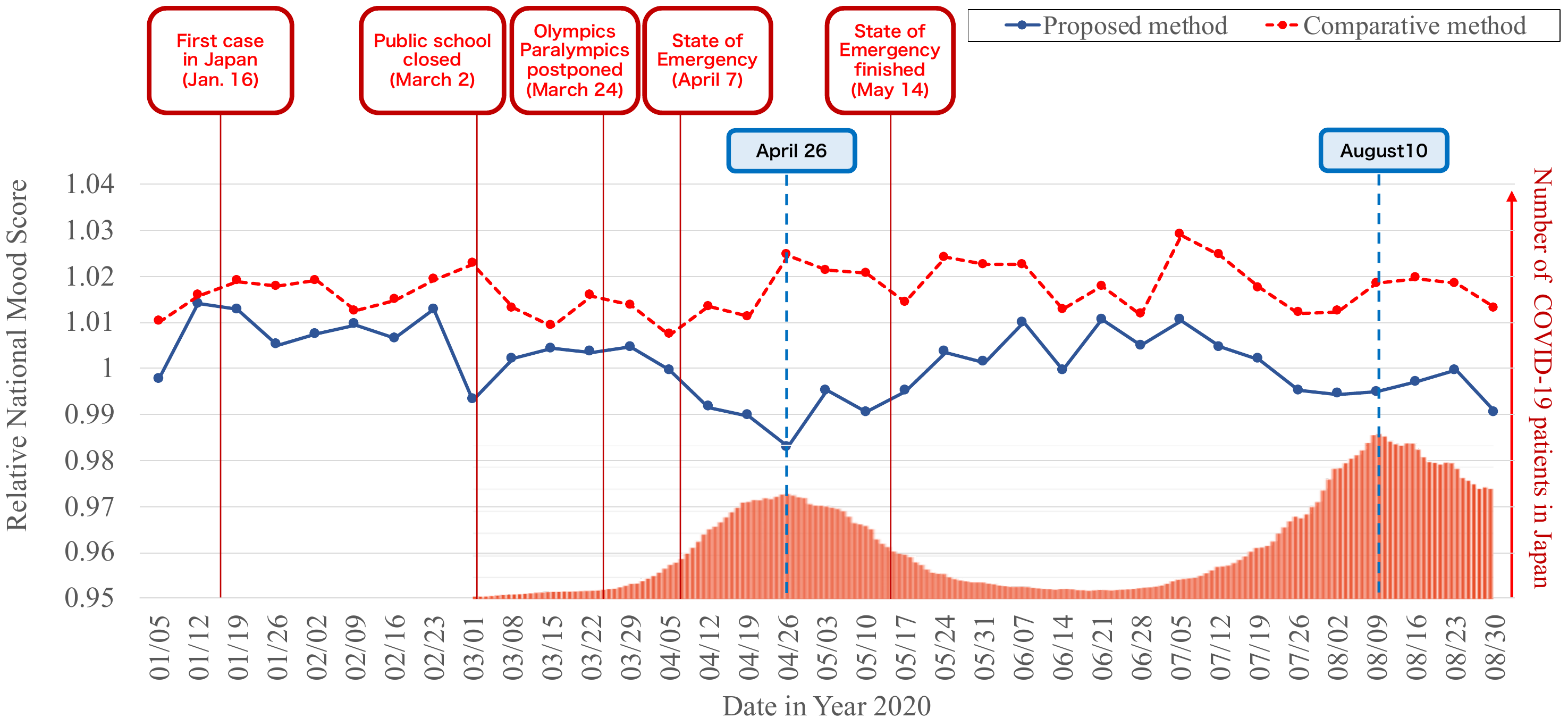}
\caption{Relative National Mood Score and Number of COVID-19 Patients in Year 2020}
\label{fig:res3}
\vspace{0.3cm}
\end{center}
\end{figure*}


Our second evaluation aims to reveal how national mood score changes in the COVID-19 pandemic situation in 2020.
In this case, we looked at the change in the daily national mood scores every Sunday from the beginning of the year to the end of July, on two different years 2018 and 2020. (The most recent stored historical data for such a long term was the data on year 2018. Due to some internal infrastructural change, we could not retrieve the equivalent data for year 2019.)
We chose Sundays since every Sunday is a holiday. 
On the other hand, in the case of other days of the week, it was assumed that the analysis results would be difficult to discuss due to occasional holidays. 

On such dataset, we firstly calculated the national mood score for each day of the period. Then, for each year, all the scores were normalized to the score of the first Sunday of the year respectively.

The result is illustrated in Figure~\ref{fig:res3}.
The x-axis shows the date. 
The lines represent the daily national mood scores. 
The value in y-axis is ``relative'' compared to the score of the same day in year 2018.
A value greater than 1 indicates a higher score in the same period in 2020 than in 2018, and vice versa.
The bar graph shows the number of COVID-19 participants in Japan in 2020~\cite{YahooJapanCOVID19summary}.

Surprisingly, the results show that a peak of the waves of COVID-19 infection spread 
and degradation of the national mood score are synchronized. 
The peak of the COVID-19 first wave was April 26 (the number of the patients: 9,577). At exactly the same day, the mood score decreased down to the bottom 0.983. 
The peak of the second wave was August 10 (the number of the patients: 15,042). The numbers around the second bottom of the score were 0.9943 (August 3) and 0.9948 (August 10). 
On the other hand, we can not confirm such clear tendency in the comparative method. 

It is very interesting to read the tendency for mood to become more negative as the number of COVID-19 patients increases, and more positive as the number of patients begins to decrease.
In 2020, Japan had originally planned to host the Tokyo Olympics and Paralympics. The mood in Japan was positive in the beginning of the year. However, COVID-19 began to rage, the Olympics were postponed, and various economic activities were restricted.
The number of corona cases began to rise in Japan, and the public was frightened of it.
Then, as the first wave subsided, the mood was positive for the restoration of economic activity again.
However, when the second wave starts to occur, the mood goes negative again.
This graph looks to have successfully tracked the tumultuous changes in Japanese people's mood in 2020.

The most remarkable point on our model (with the proposed model) is that the period of time when we collected sensor data and search queries for the model building was from October to December of 2019, which is before the COVID-19 pandemic.
This means that there is no possibility that the search queries in the trained model contain rules, such as ``A query COVID-19 is a negative feature for mood estimation''. 
Again, there is no ground truth on ``national mood''. 
However, from the fact that we can nevertheless observe the score trend inversely correlated with the number of COVID-19 patients, we conclude that the national mood score with the proposed method matches our intuition more. 


\section{Limitation and Future Work}
\label{sec:fw}
The next step in this study is the evaluation of the user’s sentiment based on real services. Our plan is to develop an actual service that can differentiate advertisements and recommendations based on mood scores. However, there are three problems with this approach.
First, we need a method for estimating which advertisements are relevant to what type of mood in advance.
Once we build such a methodology, it will be possible to evaluate the actual advertisements.
Another task is to improve the model performance. 
In this study, we adopted a white-box model to examine the effectiveness of the model qualitatively. 
We hope to employ models focused on precision and recall performance 
towards further performance improvement when we conduct real-world tests.
Finally, classification other affective statuses (beyond mood) should be possible. 
We expect that the same framework can be used to model a wide range of affective statuses. 
Our future work includes building such models and extensive evaluation of them on our services. 


\section{Conclusion}
\label{sec:Conclusion}
Affection-awareness is one of the key components in the human-centric information service. However, particularly in the real-world web field, estimating such statuses of the user is yet to be realized. We proposed a novel way estimation methodology of web user's mood based a combinational use of their search queries and mobile sensor data. Our extensive data analysis revealed multiple interesting results, including the existence of advertisements to which mood-status-based delivery would be significantly effective, and the changes of national mood scores in the weekly rhythm and in the COVID-19 pandemic situation.



\bibliographystyle{ACM-Reference-Format}
\bibliography{Okoshi_YahooEmo_Arxiv}

\end{document}